\documentclass[aps,draft,prb]{revtex4}

\begin{document}
\draft
\title{ A method for direct observation of quantum tunneling in a single molecule.}
\author{Anatoly Yu. Smirnov}
\email{anatoly@dwavesys.com}

\affiliation{ D-Wave Systems Inc. 320-1985 West Broadway,\\
 Vancouver, B.C. V6J 4Y3, Canada   }

\date{\today }

\begin{abstract}
{ An application of impedance measurement technique (IMT) for a detection of quantum tunneling in molecular structures is investigated. 
A charged particle which tunnels in a two-well potential is electrically coupled to a high-quality superconducting LC-circuit(tank) that makes possible a measurement of the electric susceptibility of the molecule at the resonant frequency of the tank. The real part of this susceptibility bears information about the tunneling rate through a measurable parameter - a phase angle between the tank voltage and a bias current applied to the tank. It is shown that the present approach is highly sensitive and allows the monitoring of the tunnel motion of charged nuclei in a single molecule.  
}
\end{abstract}

%\pacs{85.25.Cp, 03.67.Lx, 03.65.Ta, 03.65.Yz}
 \maketitle

Tunneling of a particle between minima of a double-well potential represents a pure quantum phenomenon which takes place in many physical and chemical systems \cite{Bender1,McMahon1}. Splittings of molecular energy levels due to tunneling motion of constituent nuclei have been observed  using spectroscopical methods \cite{Oppen1,Frost1,Sekiya1}
 as well as  methods of nuclear magnetic resonance relaxometry  \cite{Horsewill1}. The measurements have been performed with samples containing a large number of molecules. Recent progress in molecular manipulation with a scanning tunneling microscope tip \cite{Wolkow1,Hla1} makes possible a creation of single-molecule devices, and, in particular, apparatuses which use tunneling effects \cite{QC-NH3}. 

Here we propose to employ an impedance measurement technique \cite{Greenberg2002,APL_IMT} for contactless characterization of quantum tunneling in single molecules. This technique has provided a simple and reliable method for experimental investigation of tunneling in magnetic systems, especially, in superconducting flux qubits \cite{Izmalkov1,Grajcar1,Izmalkov2}. 
In the framework of IMT method the qubit's loop is inductively coupled to a high-quality LC circuit (tank) driven by a time-dependent bias current. An interaction of the tank with the qubit modifies an effective inductance of the tank and produces a shift of its resonant frequency. As a result, a voltage in the LC-circuit, $\langle V_T(t)\rangle = V_T \cos(\omega t + \Theta),$ has been displaced in phase from the bias current, $I_{bias}(t)=I_{ac}\cos \omega t,$ by the angle $\Theta.$ 
A dependence of the angle $\Theta$ on the bias applied to the qubit has demonstrated IMT dips which are indicative of quantum tunneling in the qubit. A tunneling rate between wells is  determined by the width of the dips. From the theoretical point of view the tangent of the 
difference between phases of voltage and current, $\tan\Theta,$ is proportional to the real part of a magnetic susceptibility of the qubit. 

To study intra-molecular tunneling of a charged particle, say, a proton or another light nucleus, in a two-well potential $U(z)$, we have to measure an electric susceptibility of such a molecule. To do this, the molecule is placed between plates of the tank's capacitor $C_T,$ so that the charged particle tunnels along lines of the intra-capacitor electric field $E_z = V_T/d.$  Here $V_T$ is a voltage applied to capacitor's plates which are separated by the distance $d$. We assume that the two-well potential $U(z)$ has minima at $z=\pm z_0.$ Taking into account the lowest tunneling doublet of energy eigenstates we describe quantum dynamics of the system by the Hamiltonian
\begin{equation}\label{eq_H}
H = \frac{\Delta}{2} \sigma_x + \frac{\varepsilon}{2} \sigma_z - (\lambda V_T + Q_0 + f) \sigma_z + H_T.
\end{equation}
Here $\Delta$ is a rate related to the particle's tunneling between the wells, $\varepsilon$ is a bias describing an asymmetry of the wells. This bias can be changed by applying an additional constant voltage to the capacitor $C_T.$ A position of the particle is characterized by the Pauli matrix $\sigma_z: \hat{z} = z_0 \sigma_z,$ so that an interaction with the intra-capacitor electric field $E_z$ is given by the term $ -\lambda V_T\sigma_z,$ where $\lambda = e(z_0/d).$ 
We add a qubit's internal heat bath  with a variable $Q_0$ and 
an auxiliary external force $f$. It should be noted that thermal fluctuations of the tank voltage contribute to decoherence and relaxation of the qubit together with internal mechanisms. 
We consider the tank, having  a resonant frequency $\omega_T = 1/\sqrt{L_TC_T},$ 
 as a quantum harmonic oscillator with the Hamiltonian ($\hbar = 1$): 
\begin{equation}\label{eq_H_T}
H_T = \omega_T(a^+a +1/2) - (a + a^+)Q_b  - L_T \hat{I}_T I_{bias} + H_{TB}.
\end{equation}
The tank's excitations are described by creation/annihilation operators $a^+,a ([a,a^+]_- =1);$ 
 in so doing for the operators of voltage and current we obtain the expressions: $\hat{V}_T = i\sqrt{\omega_T/2C_T} (a^{+} - a), \hat{I}_T = \sqrt{\omega_T/2L_T} (a^{+} + a).$ 
A linewidth broadening $\gamma_T$ and a finite quality factor of the tank, $Q_T = \omega_T /(2\gamma_T),$ are caused by the tank's own heat bath, which is  characterized by the variable $Q_b$ and the Hamiltonian $H_{TB}$. 

Following to Ref.15 
we derive the equation for the averaged voltage in the tank coupled to the double dot:
\begin{equation}\label{eq_V_T}
\left( \frac{d^2}{dt^2} + \gamma_T \frac{d}{dt} + \omega_T^2\right) \langle \hat{V}_T\rangle  = \lambda \frac{\omega_T^2}{C_T} \langle \sigma_z\rangle + \frac{1}{C_T} \dot{I}_{bias}.
\end{equation}
The particle's variable $ \langle \sigma_z\rangle $ is functionally-dependent on the tank's voltage: 
\begin{equation}\label{eq_sigma_z}
\langle \sigma_z(t) \rangle = \lambda \int dt_1 \langle \frac{\delta \sigma_z(t)}{\delta f(t_1)}\rangle \langle V_T(t_1)\rangle,
\end{equation}
where the functional derivative corresponds to the electric susceptibility $\chi_z(\omega)$ of the  particle in the double-dot system
\begin{equation}\label{eq_chi_z1}
\langle \frac{\delta \sigma_z(t)}{\delta f(t_1)}\rangle = \int \frac{d\omega}{2\pi}e^{-i\omega(t-t_1) } \chi_z(\omega).
\end{equation}
The bias current, $I_{bias}(t)=I_{ac}\cos \omega t,$ 
 applied to the tank, induces oscillations of the tank's voltage, $\langle V_T(t)\rangle = V_T \cos(\omega t + \Theta),$ with the amplitude $V_T$ and the phase $\Theta$ which obey the equation
\begin{equation}\label{eq_V_T_Theta}
\left[- \omega^2 - i \gamma_T \omega + \omega_T^2 - \frac{\lambda^2\omega_T^2}{C_T}\chi_z(\omega)\right] V_T e^{-i\Theta} = - \frac{i\omega}{C_T}I_{ac}.
\end{equation}
For the case when the frequency of the bias current is exactly equal to the initial resonant frequency of the tank, $\omega = \omega_T,$ the current-voltage phase shift is given by the equation
\begin{equation}\label{eq_tan_theta1}
\tan\Theta = - 2\lambda^2 \frac{ \bar{Q}_T}{C_T}\chi'_z(\omega_T),
\end{equation}
where $\bar{Q}_T = \omega_T/(2\bar{\gamma}_T)$ is an effective quality factor of the tank, modified because of the double dot contribution to the broadening of the tank's linewidth:   
$ \bar{\gamma}_T = \gamma_T + (\lambda^2 \omega_T/C_T) \chi''_z(\omega_T).$

The electric susceptibility of the double dot, $\chi_z(\omega),$ can be found in the framework of the theory of open quantum systems \cite{ES1981,ASRabi2003}. 
Quantum dynamics of the charged particle in two-well potential is governed by the Heisenberg equations:
\begin{eqnarray}\label{eq_Heis}
\dot{\sigma}_x = - \varepsilon \sigma_y + 2 ( Q + f) \sigma_y, \nonumber\\
\dot{\sigma}_y = - \Delta \sigma_z +  \varepsilon \sigma_x - 2 ( Q + f) \sigma_x, \nonumber\\
\dot{\sigma}_z = \Delta \sigma_y.
\end{eqnarray}
Here $Q = Q_0 + \lambda \tilde V_T $ is the operator of the total dissipative environment which surrounds the double dot. For the susceptibility of this environment we have the relation: $\chi(\omega) = \chi_0(\omega) + \chi_T(\omega),$ where $\chi_0(\omega)$ corresponds to the internal mechanisms of dissipation in the double dot, and 
\begin{equation}\label{chi_T}
\chi_T (\omega) = \frac{e^2}{C_T} \left(\frac{z_0}{d} \right)^2 \frac{\omega_T^2}{\omega_T^2 - \omega^2 - i \omega \gamma_T}
\end{equation}
describes dissipative properties of the tank which are due to its coupling to the bath $Q_b.$
The spectral functions of the heat bath fluctuations, $S(\omega) = S_0(\omega) + S_T(\omega),$ 
is related to the imaginary part of the heat bath susceptibility by the fluctuation-dissipation theorem: $ S(\omega ) =  \chi''(\omega) \coth(\omega/2T),$ with $T$ being a temperature of the bath. Averaging the Heisenberg equations (\ref{eq_Heis}) over the equilibrium fluctuations of environment followed by an application of Bloch-Redfield approximation allows us to find the electric susceptibility of the charged particle confined in the two-well potential
\begin{equation}\label{eq_chi_z2}
\chi_z(\omega) = - 2 \Delta \frac{- i \omega + S(\omega + \omega_0) + S(\omega - \omega_0)}
{(-i \omega + \gamma_0)[ - i (\omega - \omega_0) + \gamma] [ - i(\omega+ \omega_0) + \gamma] }   \sigma_x^0.
\end{equation}
Here $\omega_0 = \sqrt{\Delta^2 + \varepsilon^2}$ is the ground-state splitting for the tunneling particle, $\sigma_x^0 = - (\Delta/\omega_0)\tanh(\omega_0/2T)$ is a steady-state value of the matrix $\sigma_x.$ Coupling of the double dot to the dissipative environment results in the frequency-dependent decoherence and relaxation rates $\gamma$ and $\gamma_0$
\begin{eqnarray}\label{eq_gamma1}
\gamma(\omega) = \frac{\Delta^2}{\omega_0^2} S(\omega) + 
\frac{\varepsilon^2}{\omega_0^2}\left(1 - \frac{\omega_0}{\omega} \right)  S(\omega+\omega_0) + 
\frac{\varepsilon^2}{\omega_0^2}\left(1 + \frac{\omega_0}{\omega} \right)  S(\omega-\omega_0), 
\nonumber\\
\gamma_0(\omega) =  \frac{\Delta^2}{\omega_0^2} \left[ S(\omega+\omega_0) + S(\omega - \omega_0) \right].
\end{eqnarray}
In particular, a back-action of the tank on the tunneling particle leads to the small decoherence and relaxation rates 
\begin{eqnarray}\label{eq_gamma}
\gamma(\omega_0)= \frac{1}{2} \gamma_0(0) = \left(\frac{\Delta}{\omega_0}\right)^2 S_T(\omega_0) =  \frac{e^2}{2 Q_T C_T}\left(\frac{z_0}{d}\right)^2 \left(\frac{\Delta}{\omega_0}\right)^2 \left(\frac{\omega_T}{\omega_0}\right)^3.
\end{eqnarray}
It should be emphasized that the current-voltage angle $\Theta$ (\ref{eq_tan_theta1}) is proportional to the double-dot susceptibility, $\chi'_z(\omega_T),$ taken at the resonant frequency of the tank $\omega_T, $ which is much lower than the tunneling frequency  $\omega_0.$ 
Taking this fact into account we find the real part of the electric susceptibility: $\chi'_z(\omega_T) = - 2 \Delta\sigma_x^0/\omega_0^2. $ 
As a result, the phase angle between the voltage in the tank and the bias current is determined  by the following equation
\begin{equation}\label{eq_tan2}
\tan\Theta = - \frac{e^2}{C_T} \left( \frac{2z_0}{d}\right)^2 Q_T \frac{\Delta^2}{(\Delta^2 + \varepsilon^2)^{3/2}} \tanh\left( \frac{\sqrt{\Delta^2 + \varepsilon^2}}{2T}\right).
\end{equation}

By way of illustration, we consider  tunneling in an ammonia molecule NH$_3$  which is described by the two-well potential energy \cite{FLP}. The existence of two delocalized states with eigenenergies separated by a tunneling splitting $\Delta/h = 23.8 $ GHz  has motivated   to harness this molecule as a quantum bit \cite{QC-NH3}.  According to this proposal the NH$_3$-molecule can be encapsulated into a fullerene $C_{60}$ to be addressed individually. A direct chemisorption of the ammonia molecule on the surface of a semiconductor represents another possiblity \cite{Brown1}.
The fullerene ball incorporating the single ammonia molecule is inserted between the capacitor's plates having an area about 1 $\mu m^2$. If the distance between plates is of order of a diameter of $C_{60}-$molecule \cite{ParkC60}: $d =1.2 \times 10^{-9} m $ , then for the capacitance $C_T$ we get an estimation: $C_T \simeq  $ 10 fF.  Together with a value for the tank's inductance, $L_T \simeq $ 1 $ \mu H,$ it gives the resonant frequency of the tank: 
$\omega_T/2\pi = $ 1.6 GHz which is much less than the tunneling frequency $\Delta/h: \hbar \omega_T/\Delta = 0.06. $ The electric dipole moment of the ammonia molecule \cite{Brown1}, $2ez_0,$ which is involved in  the formula (\ref{eq_tan2}), is of order of  0.31  $e \times 10^{-10}$  m . For the case when losses in the tank are mainly due to a shunt resistance $R_T,$ which is parallel to the capacitor $C_T$, the linewidth of the tank is defined as: $\gamma_T = 1/(R_TC_T), $ so that the quality factor is $Q_T = \omega_T R_T C_T/2,$  and 
the current-voltage angle $\Theta$ does not depend on the capacitance of the tank $C_T:$
\begin{equation}\label{eq_tan3}
\tan\Theta = - \frac{\hbar \omega_T}{2\Delta} \frac{R_T}{(\hbar/e^2)} \left( \frac{2z_0}{d}\right)^2 
 \frac{\Delta^3}{(\Delta^2 + \varepsilon^2)^{3/2}} \tanh\left( \frac{\sqrt{\Delta^2 + \varepsilon^2}}{2T}\right).
\end{equation}
The current-voltage angle $\Theta$ can be measured in the experiment as a function of the bias $\varepsilon$ applied to the tunneling particle. With a reasonable estimation for the shunt resistance \cite{ParkC60}: 
$ R_T= 50 \times 10^6 $ Ohm, the single ammonia molecule demonstrates a pronounced low-frequency response: $\tan \Theta = - 0.23,$ at temperatures $T < \Delta/k_B = $ 1.1 K, and $\varepsilon = 0.$
Measurement-induced decoherence, $\gamma = \gamma(\Delta)$ (\ref{eq_gamma}), is negligibly small in the process: $\gamma /\Delta \simeq 10^{-11},$ that allows us to characterize the IMT procedure as a weak continuous measurement \cite{Averin1,Korotkov1}. 
To be realistic, for the example under discussion - the ammonia molecule confined inside of $C_{60}$  - we have to take into account screening effects due to a fullerene cage \cite{QC-NH3}. Because of the screening, a magnitude of the IMT dip 
(\ref{eq_tan2},\ref{eq_tan3}) decreases that can be compensated by a proper increasing of the shunt resistance $R_T$ and the quality factor of the tank $Q_T$.  

In conclusion, we have proposed to use the impedance measurement technique for continuous monitoring of quantum tunneling in molecular systems which are characterized by a two-well potential energy. To this end, the molecule(s) should be inserted between the plates of a capacitor $C_T$ comprising a high-quality superconducting tank along with an inductance $L_T$. We have shown that the present approach is sensitive enough to detect a tunneling motion in a single molecule. 

The author is grateful to M. Grajcar, E. Il'ichev, A. Izmalkov, S. Rashkeev, M. Steininger, and A. Zagoskin for helpful comments.

\end{document}